\documentclass{aa}
\usepackage{graphicx}
\usepackage[varg]{txfonts}
\usepackage{cases}
\usepackage{natbib}

\begin{document}
   \title{Energy partition in a confined flare with an extreme-ultraviolet late phase}

   \author{Q. M. Zhang\inst{1,3}, J. X. Cheng\inst{2}, Y. Dai\inst{4}, K. V. Tam\inst{5}, and A. A. Xu\inst{5}}

   \institute{Key Laboratory of Dark Matter and Space Astronomy, Purple Mountain Observatory, CAS, Nanjing 210023, PR China \\
              \email{zhangqm@pmo.ac.cn}
              \and
              Key Laboratory of Planetary Sciences, Shanghai Astronomical Observatory, Shanghai 200030, PR China \\
              \and
              CAS Key Laboratory of Solar Activity, National Astronomical Observatories, CAS, Beijing 100101, PR China \\
              \and
              School of Astronomy and Space Science, Nanjing University, Nanjing 210023, PR China \\
              \and
              State Key Laboratory of Lunar and Planetary Sciences, Macau University of Science and Technology, Macau, PR China \\
              }

   \date{Received; accepted}
    \titlerunning{Energy partition in a confined flare}
    \authorrunning{Zhang et al.}

 \abstract
   {}
   {In this paper, we reanalyze the M1.2 confined flare with a large extreme-ultraviolet (EUV) late phase on 2011 September 9, focusing on its energy partition.}
   {The flare was observed by the Atmospheric Imaging Assembly (AIA) on board the Solar Dynamics Observatory (SDO).
   The three-dimensional (3D) magnetic fields of active region 11283 before flare were obtained using the nonlinear force free field modeling and the vector magnetograms 
   observed by the Helioseismic and Magnetic Imager (HMI) on board SDO. Properties of the nonthermal electrons injected into the chromosphere were obtained from the hard X-ray
   observations of the Ramaty Hight Energy Solar Spectroscopic Imager (RHESSI). Soft X-ray fluxes of the flare were recorded by the GOES spacecraft.
   Irradiance in 1$-$70 {\AA} and 70$-$370 {\AA} were measured by the EUV Variability Experiment (EVE) on board SDO. Various energy components of the flare are calculated.}
   {The radiation ($\sim$5.4$\times$10$^{30}$ erg) in 1$-$70 {\AA} is nearly eleven times larger than the radiation in 70$-$370 {\AA}, 
   and is nearly 180 times larger than the radiation in 1$-$8 {\AA}.
   The peak thermal energy of the post-flare loops is estimated to be (1.7$-$1.8)$\times$10$^{30}$ erg based on a simplified schematic cartoon.
   Based on previous results of Enthalpy-Based Thermal Evolution of Loops (EBTEL) simulation, 
   the energy inputs in the main flaring loops and late-phase loops are (1.5$-$3.8)$\times$10$^{29}$ erg and 7.7$\times$10$^{29}$ erg, respectively.
   The nonthermal energy ((1.7$-$2.2)$\times$10$^{30}$ erg) of the flare-accelerated electrons is comparable to the peak thermal energy 
   and is sufficient to provide the energy input of the main flaring loops and late-phase loops. 
   The magnetic free energy (9.1$\times$10$^{31}$ erg) before flare is large enough to provide the heating requirement and radiation, 
   indicating that the magnetic free energy is adequate to power the flare.}
   {}


 \maketitle

\section{Introduction} \label{s-intro}
Solar flares and coronal mass ejections (CMEs) are the most powerful activities in the solar atmosphere and have potential influences on space weather \citep{for06,fle11,rea15,pats20}. 
A total amount of 10$^{29}$-10$^{32}$ erg magnetic free energy is impulsively released and converted into various kinds of energy, including the thermal energy of the 
directly heated plasma, kinetic energy of the bulk outflows, nonthermal energies of accelerated electrons and ions, and kinetic energy of a CME \citep[e.g.,][]{pri02,sto07,sch08,jing09,ems12}.
According to their association with CMEs, flares can be classified into eruptive events and confined events \citep{moo01}.
Confined flares are produced by loop-loop interaction \citep{mot20} or triggered by failed filament eruptions \citep{ji03,song14,yan20}.
The strong strapping field overlying the active region (AR) core plays a dominant role in constraining the eruptions \citep{wang07,my15,sun15,ama18,bau18}.

Energy partition in flares is an important issue, which has been extensively explored \citep[e.g.,][]{sai02,ems04,ems05,ems12,cham12,feng13,han13,mil14,war16a,war16b,war20,mot20}. 
\citet{sto07} studied 18 microflares observed by the Ramaty Hight Energy Solar Spectroscopic Imager \citep[RHESSI;][]{lin02}. 
The average thermal and nonthermal energies are 7$\times$10$^{27}$ erg and 2$\times$10$^{29}$ erg, respectively.
\citet{ing14} investigated the energetics of ten microflares. It is concluded that multi-thermal plasma is a relevant consideration for microflares.
For large eruptive flares, the energies contained in flares and CMEs are roughly comparable \citep{ems05,ree10,feng13}.
The global energetics of nearly 400 flares and CMEs during 2010$-$2013 were studied in a comprehensive way \citep{asch15,asch17}. 
It is found that the nonthermal energy exceeds the thermal energy in 85\% of the whole events, which is consistent with the collisional thick-target model \citep{bro71}. 
So far, few work has been dedicated to the energy partition in confined flares.
\citet{tha15} studied the X1.6 flare on 2014 October 22 and calculated the nonthermal energy in electrons and magnetic free energy in AR 12192. 
The nonthermal energy accounts for 10\% of the magnetic free energy.
\citet{zqm19} investigated the energy partition in two M1.1 circular-ribbon flares (CRFs) in AR 12434 on 2015 October 15 and 16. 
The peak thermal energy, nonthermal energy of flare-accelerated electrons, total radiative loss of hot plasma, and radiative output in 1$-$8 {\AA} and 1$-$70 {\AA} of the flares are calculated.
It is revealed that the two confined flares have similar energetics and the total heating requirement of flare loops could adequately be supplied by nonthermal electrons.
Recently, \citet{cai21} studied the energetics of four confined CRFs in detail. The ratio of nonthermal energy in flare-accelerated electrons to the magnetic free energy lies in the range
0.70$-$0.76, which is much higher than that of eruptive flares. Hence, this ratio is proposed to be an essential factor for discriminating confined from eruptive flares.

Normally, the complete evolution of soft X-ray (SXR) flux of a flare is characterized by a slow rise in the preflare phase, a sharp rise in the impulsive phase, and a gradual decay phase \citep{zj01}.
Shortly after the launch of Solar Dynamics Observatory \citep[SDO;][]{pes12}, an extra peak in the extreme-ultraviolet (EUV) ``warm'' coronal lines (e.g., Fe\,{\sc xvi} 335 {\AA}, $\sim$2.5 MK)
was observed by the EUV Variability Experiment \citep[EVE;][]{wood12} on board SDO and was termed ``EUV late phase'' \citep{wood11}. The time lag between the main phase 
and late phase ranges from a few tens of minutes to several hours. The emission of late phase is found to originate from a set of coronal loops higher and longer than the 
main flaring loops \citep{wood11,dai13,liu13,sun13,li14,mas17,wang20}. In most cases, the peak of late phase is lower than the main-phase peak \citep{wood11}. 
Nevertheless, some of the events show a stronger late-phase peak than the main-phase peak, which is termed extreme EUV late phase \citep{liu15,dai18b,zhou19}.
The flares with a late phase could be associated with circular ribbons, two ribbons, or complex multiple ribbons \citep{chen20}.

Although a great number of case studies have been carried out, the nature of EUV late phase is still under debate. 
One explanation is additional heating during the gradual phase \citep{dai13,liu15,kuh17}, 
and the other is prolonged cooling time of the higher and longer late-phase loops since the conductive cooling timescale increases with the loop length \citep{car94,liu13,mas17}.
Combining observations and numerical simulations using the Enthalpy-Based Thermal Evolution of Loops \citep[EBTEL;][]{kli08,car12a,car12b} code, \citet{sun13} concluded that
the second peak comes from the long cooling of large post-reconnection loops and additional heating may also be required. That means both mechanisms may be at work \citep{li14}.
\citet{dai18a} investigated the effects of these two processes on the emission characteristics in late-phase loops. It is shown that although both processes can generate
an EUV late phase that is consistent with observations, the hydrodynamic and thermodynamic evolutions in late-phase loops are quite different for the two different processes.

In 2011 September, a series of flares occurred in AR 11283. 
Using a nonlinear force-free field (NLFFF) extrapolation prior to the filament eruption as the initial condition in a model, 
\citet{jia13} performed three-dimensional (3D) magnetohydrodynamic (MHD) simulation of the eruption on September 6. The simulation nicely reproduces the realistic initiation process. 
\citet{pra20} also carried out MHD simulation of the magnetic null-point reconnections and coronal dimmings during the X2.1 flare on September 6.
\citet{zqm15} studied the M6.7 flare on September 8. The flare was triggered by the partial eruption of the major part of a sigmoidal filament, 
while the ejection of a runaway part caused a small and weak CME.
\citet{dai18b} analyzed an M1.2 confined flare with an extremely large EUV late phase on September 9. 
Using the EBTEL code, the authors modeled the EUV emissions of a late-phase loop, obtaining a peak heating rate of 1.1 erg cm$^{-3}$ s$^{-1}$.
Till now, energy partition in a confined flare with an EUV late phase has not been studied.

In this paper, we reanalyze the M1.2 flare on 2011 September 9 and focus on the energy partition building on the work of \citet{dai18b}.
The rest of the paper is organized as follows. In Sect.~\ref{s-data}, we describe the observations and data analysis. 
The calculations of different energy components are elucidated in Sect.~\ref{s-eng}. 
We compare our findings with previous works in Sect.~\ref{s-disc} and give a brief summary in Sect.~\ref{s-sum}.

\section{Observations and data analysis} \label{s-data}
We use multiwavelength observations from SDO, GOES, and RHESSI.
The Atmospheric Imaging Assembly \citep[AIA;][]{lem12} on board SDO takes full-disk images in two UV (1600 and 1700 {\AA}) and seven EUV (94, 131, 171, 193, 211, 304, and 335 {\AA}) wavelengths. 
The photospheric line-of-sight (LOS) and vector magnetograms are observed by the Helioseismic and Magnetic Imager \citep[HMI;][]{sch12} on board SDO. 
The AIA and HMI level\_1 data are calibrated using the standard Solar Software (SSW) programs \texttt{aia\_prep.pro} and \texttt{hmi\_prep.pro}, respectively. 
The irradiance from a broad band ranging from 1$-$70 {\AA} is directly measured by the EUV SpectroPhotometer (ESP) on board EVE.
The Multiple EUV Grating Spectrographs (MEGS)-A on board EVE, covering the 6$-$37 nm range, records a complete spectrum with a time cadence of 10 s and a spectra resolution of 1 {\AA}.
The standard SSW program \texttt{eve\_integrate\_line.pro} is employed to integrate irradiance over 70$-$370 {\AA} using the EVS spectral data from MEGS-A.
SXR light curves of the flare in 0.5$-$4 {\AA} and 1$-$8 {\AA} are recorded by the GOES spacecraft. The isothermal temperature ($T_{e}$) and emission measure (EM) of the 
SXR-emitting plasma are derived from the ratio of GOES fluxes \citep{wh05}.

To derive the background-subtracted HXR spectra observed by RHESSI, the pulse pileup correction, energy gain correction, and isotropic albedo photo correction are conducted. 
To obtain the properties of flare-accelerated electrons, 
we fit the HXR spectra by the combination of an isothermal component determined by the isothermal temperature and EM and a nonthermal component created by
the thick-target bremsstrahlung of energetic electrons with a low-energy cutoff \citep{war16a}. The spectra fitting is conducted 
using the OSPEX\footnote{https://hesperia.gsfc.nasa.gov/ssw/packages/spex/doc/ospex\_explanation.htm} software built in SSW.
The observational parameters are summarized in Table~\ref{tab-1}.

\begin{table}
\centering
\caption{Description of the observational parameters.}
\label{tab-1}
\begin{tabular}{cccc}
\hline\hline
Instrument & $\lambda$   & Cadence & Pixel Size \\ 
                  & ({\AA})         & (s)           & (\arcsec) \\
\hline
SDO/AIA  & 131, 335 &  12  & 0.6 \\
SDO/AIA  & 1600       &  24   & 0.6 \\
SDO/HMI & 6173      &  45, 720    & 0.6 \\
SDO/EVE & 1$-$70 &  0.25 & ... \\
SDO/EVE & 70$-$370 & 10 & ... \\
GOES     & 0.5$-$4 &  2.05 & ... \\
GOES     & 1$-$8    &  2.05 & ... \\
RHESSI   & 3$-$100 keV &  4.0 & 2.0 \\
\hline
\end{tabular}
\end{table}

\section{Energy partition} \label{s-eng}
\subsection{Overview of the event} \label{s-flare}
As described in \citet{dai18b}, the confined M1.2 flare in AR 11283 (N16W56) underwent two-stage energy release in the impulsive phase.
The first stage, peaking at $\sim$12:42:30 UT, 
resulted from magnetic reconnection at the null point and the surrounding quasi-sepatratrix layer \citep[QSL;][]{dem96} after the underlying flux rope rose up.  
The second stage, peaking at $\sim$12:47:30 UT, resulted from magnetic reconnection between the two legs of the field lines stretched by the flux rope that underwent a failed eruption. 
The first stage is markedly more impulsive than the second one.

\begin{figure}
\includegraphics[width=9cm,clip=]{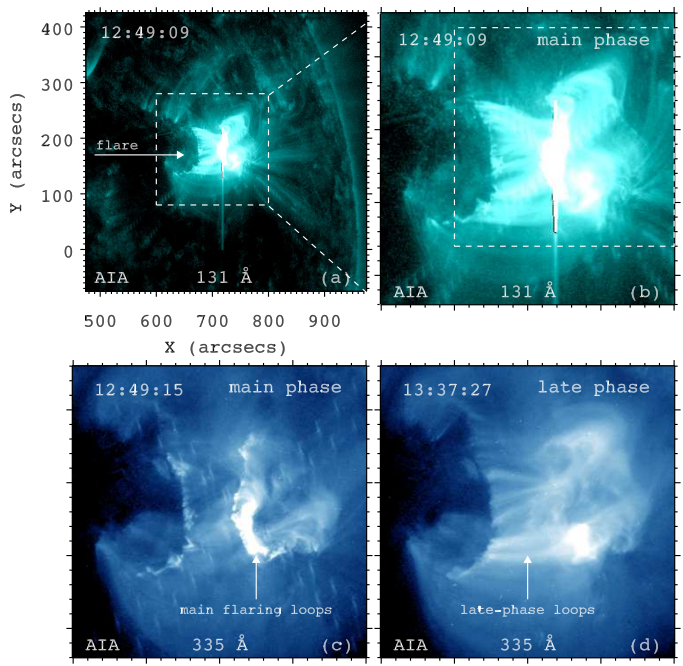}
\centering
\caption{(a) AIA 131 {\AA} image at 12:49:09 UT. The white dashed box delineates the flare region. The arrow points to the M1.2 flare. (b) Closeup of the flare region.
(c)-(d) AIA 335 {\AA} images at 12:49:15 UT and 13:37:27 UT. The arrows point to the short main flaring loops and long late-phase loops.
The whole evolution of the event observed in 131 {\AA} and 335 {\AA} is shown in a movie (\textit{flare.mp4}) available online.}
\label{fig1}
\end{figure}

In Fig.~\ref{fig1}, the AIA 131 {\AA} image observed at 12:49:09 UT is shown in panel (a). The flare region composed of very hot plasma is enclosed by the white dashed box.
A closeup of the flare region is shown in panel (b). The AIA 335 {\AA} images observed at 12:49:15 UT and 13:37:27 UT, representing the main phase and EUV late phase, 
are displayed in the bottom panels. The short, compact main flaring loops and long, dispersed late-phase loops are indicated by the arrows.
\citet{dai18b} proposed that the reconnections during the first and second stages are responsible for the EUV emissions of the late-phase loops and main flaring loops, respectively.
However, the observations seem not to be in agreement with this point, since the bright main flaring loops are clearly heated in both stages.
Hence, we propose that the first-stage release is responsible for the heating of main flaring loops and late-phase loops, while the second-stage release is uniquely responsible for the 
heating of main flaring loops. In other words, the main flaring loops are heated twice, while the late-phase loops are heated once (see Fig.\,5(e) of \citet{dai18b}).

\begin{figure}
\includegraphics[width=8cm,clip=]{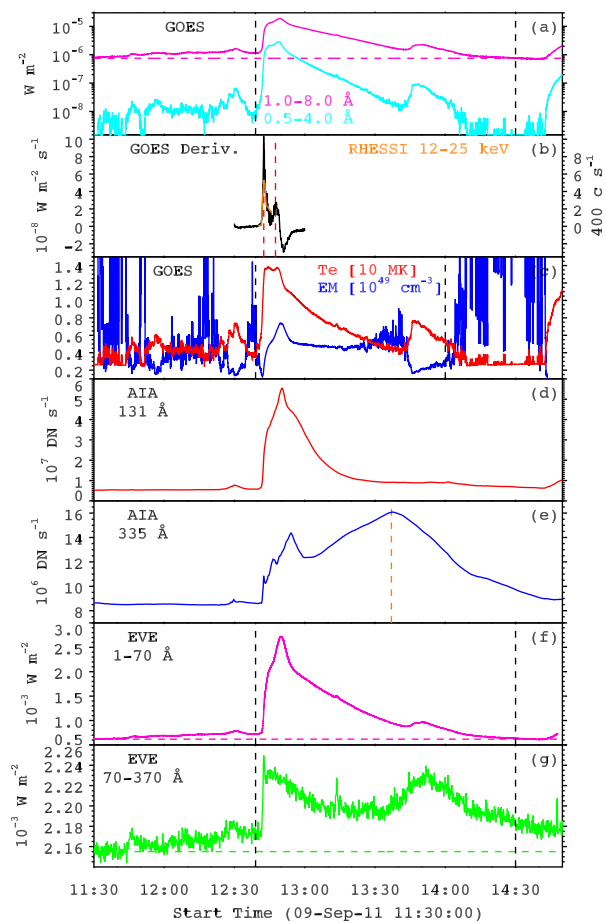}
\centering
\caption{(a) SXR light curves of the flare in 0.5$-$4 {\AA} and 1$-$8 {\AA}. The magenta dashed line represents the background intensity.
(b) Time derivative of the flux in 1$-$8 {\AA} (black line) and HXR light curve at 12$-$25 keV (orange line). 
The red dashed lines denote the peak times of two-stage energy release.
(c) Time evolutions of flare temperature ($T_{e}$) and emission measure (EM) derived from GOES observations.
(d)-(e) Time evolutions of the integral intensity of the flare in 131 {\AA} and 335 {\AA}. The orange dashed line denote the peak time of EUV late phase.
(f)-(g) Light curves of the flare in 1$-$70 and 70$-$370 {\AA}. The magenta and green dashed lines represent the background intensities, respectively.}
\label{fig2}
\end{figure}

In Fig.~\ref{fig2}, SXR light curves in 0.5$-$4.0 {\AA} and 1$-$8 {\AA} are plotted with cyan and magenta lines in panel (a). 
The SXR fluxes increase rapidly from $\sim$12:39 UT to the peak at 12:49:19 UT, before declining gradually to $\sim$14:30 UT. Hence, the lifetime of the flare reaches $\sim$110 minutes.
Time derivative of the 1$-$8 {\AA} flux during 12:30$-$13:00 UT is plotted with a black line in Fig.~\ref{fig2}(b). 
The HXR light curve at 12$-$25 keV from RHESSI observation is superposed with an orange line in Fig.~\ref{fig2}(b).
Two peaks at $\sim$12:42:30 UT and $\sim$12:47:30 UT signify the two-stage energy release (see also Fig.\,5(f) of \citet{dai18b}).
The 131 {\AA} and 335 {\AA} intensities within the white dashed box of Fig.~\ref{fig1}(b) are accumulated. 
Time evolutions of their total emissions are displayed in Fig.~\ref{fig2}(d)-(e) (see also Fig.\,1 of \citet{dai18b}).
The 131 {\AA} light curve shows a good correlation with the 1$-$8 {\AA} light curve.
The peak time of EUV late phase in 335 {\AA} is $\sim$13:37:30 UT, which has a delay of $\sim$48 minutes relative to the SXR peak time.
In the following, we focus on different energy contents of the flare.

\begin{figure}
\includegraphics[width=8cm,clip=]{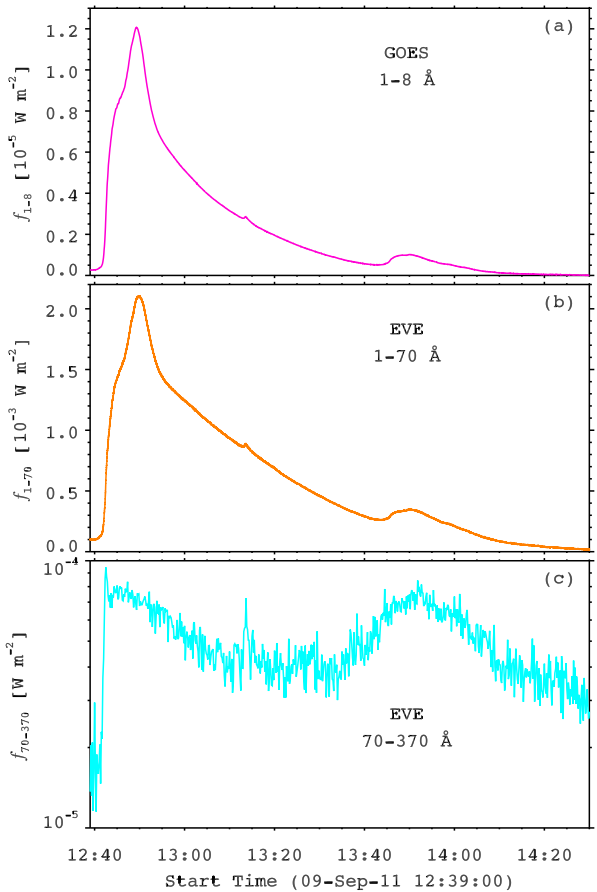}
\centering
\caption{Background-subtracted light curves of the flare in 1$-$8, 1$-$70, and 70$-$370 {\AA}.}
\label{fig3}
\end{figure}

\begin{table*}
\centering
\caption{Energy components (in units of 10$^{30}$ erg) for the M1.2 flare.}
\label{tab-2}
\begin{tabular}{cccccccc}
\hline\hline
1$-$8 {\AA} &  1$-$70 {\AA}  & 70$-$370 {\AA}  & 1$-$370 {\AA}  & $T_{rad}$ & $E_{th}$ & $E_{nth}$  & $E_{mag}$ \\ 
\hline
3.1e-2 & 5.4 & 4.7e-1 & 5.9 & 1.1 & 1.7$-$1.8 & 1.7$-$2.2 & 91 \\
\hline
\end{tabular}
\end{table*}

\subsection{Radiated energy} \label{s-rad}
First of all, we calculate the radiated energy in GOES 1$-$8 {\AA}. Note that the GOES XRS fluxes have been corrected by a factor of $\sim$1.5 \citep{wood17}.
In Fig.~\ref{fig2}(a), the flux well before the flare impulsive phase is considered as background radiation (magenta dashed line).
The background-subtracted light curve ($f_{1-8}$) is plotted in Fig.~\ref{fig3}(a).
The radiated energy is calculated by integrating $f_{1-8}$ during 12:39$-$14:30 UT after multiplying a factor of $c_0=10^7\times2\pi d^2=1.406\times10^{30}$ m$^2$, 
where $d=1.496\times10^{11}$ m represents the average distance between the Sun and Earth \citep{zqm19}.
The radiated energy in 1$-$8 {\AA} is estimated to be $\sim$3.1$\times$10$^{28}$ erg and is listed in the first column of Table~\ref{tab-2}.

Secondly, we calculate the radiated energy in EVE 1$-$70 {\AA}.
The 1$-$70 {\AA} light curve of the flare is plotted in Fig.~\ref{fig2}(f). 
Like in 1$-$8 {\AA}, the flux well before the flare impulsive phase is considered as background radiation (magenta dashed line).
The background-subtracted light curve ($f_{1-70}$) is plotted in Fig.~\ref{fig3}(b).
The radiated energy is calculated by integrating $f_{1-70}$ during 12:39$-$14:30 UT after multiplying $c_0$.
The radiated energy in 1$-$70 {\AA} is estimated to be $\sim$5.4$\times$10$^{30}$ erg, which is $\sim$180 times larger than that in 1$-$8 {\AA}.

Finally, we calculate the radiated energy in EVE 70$-$370 {\AA}.
The 70$-$370 {\AA} light curve of the flare is plotted in Fig.~\ref{fig2}(g).
Likewise, the flux well before the flare impulsive phase is considered as background radiation (green dashed line).
The background-subtracted light curve ($f_{70-370}$) is plotted in Fig.~\ref{fig3}(c).
The radiated energy is calculated by integrating $f_{70-370}$ during 12:39$-$14:30 UT after multiplying $c_0$.
The calculated radiated energy (4.7$\times$10$^{29}$ erg) in 70$-$370 {\AA} is one order of magnitude lower than the radiative output in 1$-$70 {\AA}.
Hence, the total radiated energy in 1$-$370 {\AA} reaches $\sim$5.9$\times$10$^{30}$ erg.

\subsection{Radiative loss from the SXR-emitting plasma} \label{s-radloss}
As described in \citet{feng13}, the total optically thin radiative loss from the SXR-emitting plasma is calculated based on EM and $T_{e}$ derived from GOES observations, 
which are drawn with blue and red lines in Fig.~\ref{fig2}(c).
\begin{equation} \label{eqn1}
  T_{rad}=\int_{t_1}^{t_2}\mathrm{EM}(t)\times\Lambda(T_{e}(t))dt,
\end{equation}
where $\Lambda(T_{e})$ denotes the radiative loss function \citep{cox69}, which is shown in Fig. 8 of \citet{zqm19}.
The start and end times for the integral are set to be 12:39 UT and 14:00 UT, since the signals after 14:00 UT are chaotic and unconvincing.
The value of $T_{rad}$ is estimated to be $\sim$1.1$\times$10$^{30}$ erg and is listed in the sixth column of Table~\ref{tab-2}.
$T_{rad}$ is $\sim$36 times larger than the radiation in 1$-$8 {\AA}, which is in line with previous results for confined flares \citep{zqm19,cai21}.

\subsection{Peak thermal energy of the SXR-emitting plasma} \label{s-th}
The thermal energy of the SXR-emitting plasma is:
\begin{equation} \label{eqn2}
  E_{th}=3n_{e}k_{B}T_{e}fV=3k_{B}T_{e}\sqrt{\mathrm{EM}\times fV},
\end{equation}
where $n_{e}=\sqrt{\mathrm{EM}/V}$ is the electron number density, $k_B=1.38\times10^{-16}$ erg K$^{-1}$ is the Boltzmann constant, $V$ is the total volume of the hot plasma, 
and $f\approx1$ represents the volumetric filling factor \citep{sto07,ems12}.

\begin{figure}
\includegraphics[width=9cm,clip=]{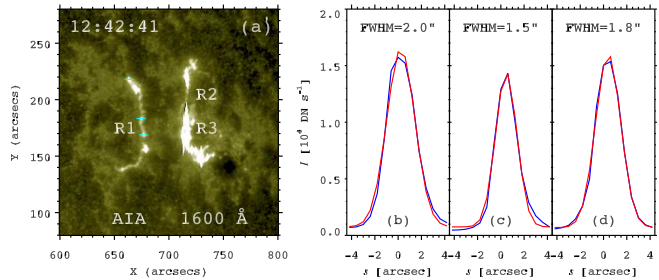}
\centering
\caption{(a) AIA 1600 {\AA} image at 12:42:41 UT. R1, R2, and R3 signify the three flare ribbons. Three representative points (green lines) are selected.
(b)-(d) Intensity distributions across the ribbon (blue lines) and results of Gaussian fitting (red lines). The values of FWHM are labeled.}
\label{fig4}
\end{figure}

Despite the flare region in Fig.~\ref{fig1}(b) is large, we note that a significant proportion of the flare region is not filled with hot plasma in 3D, which is indicated in Fig.\,7(c) of \citet{dai18b}.
Therefore, the volume of the SXR-emitting plasma is taken to be the total volume of magnetic flux tubes rooted in the three flare ribbons (R1, R2, and R3) in the chromosphere.
In Fig.~\ref{fig4}, the left panel shows the AIA 1600 {\AA} image at 12:42:41 UT. Two parallel ribbons (R1 and R2) and one short ribbon (R3) very close to R2 are obviously demonstrated.
Three representative points (green lines) on R1 are selected.
In panels (b)-(d), the intensity distributions across the ribbon are displayed with blue lines and the results of Gaussian fitting are displayed with red lines.
The full width at half maximum (FWHM) of the Gauss function lies in the range from 1$\farcs$5 to 2$\arcsec$. The mean value (1$\farcs$8) is taken to be the width of flare ribbons.
The area of R1 can be expressed as $A_{\rm R1}=l_{\rm R1}\times w$, where $l_{\rm R1}=115\arcsec$ and $w=1\farcs8$ denote the length and width of R1.
Likewise, the area of R2 is expressed as $A_{\rm R2}=l_{\rm R2}\times w$, where $l_{\rm R2}=130\arcsec$ and $w=1\farcs8$ denote the length and width of R2.

\begin{figure}
\includegraphics[width=9cm,clip=]{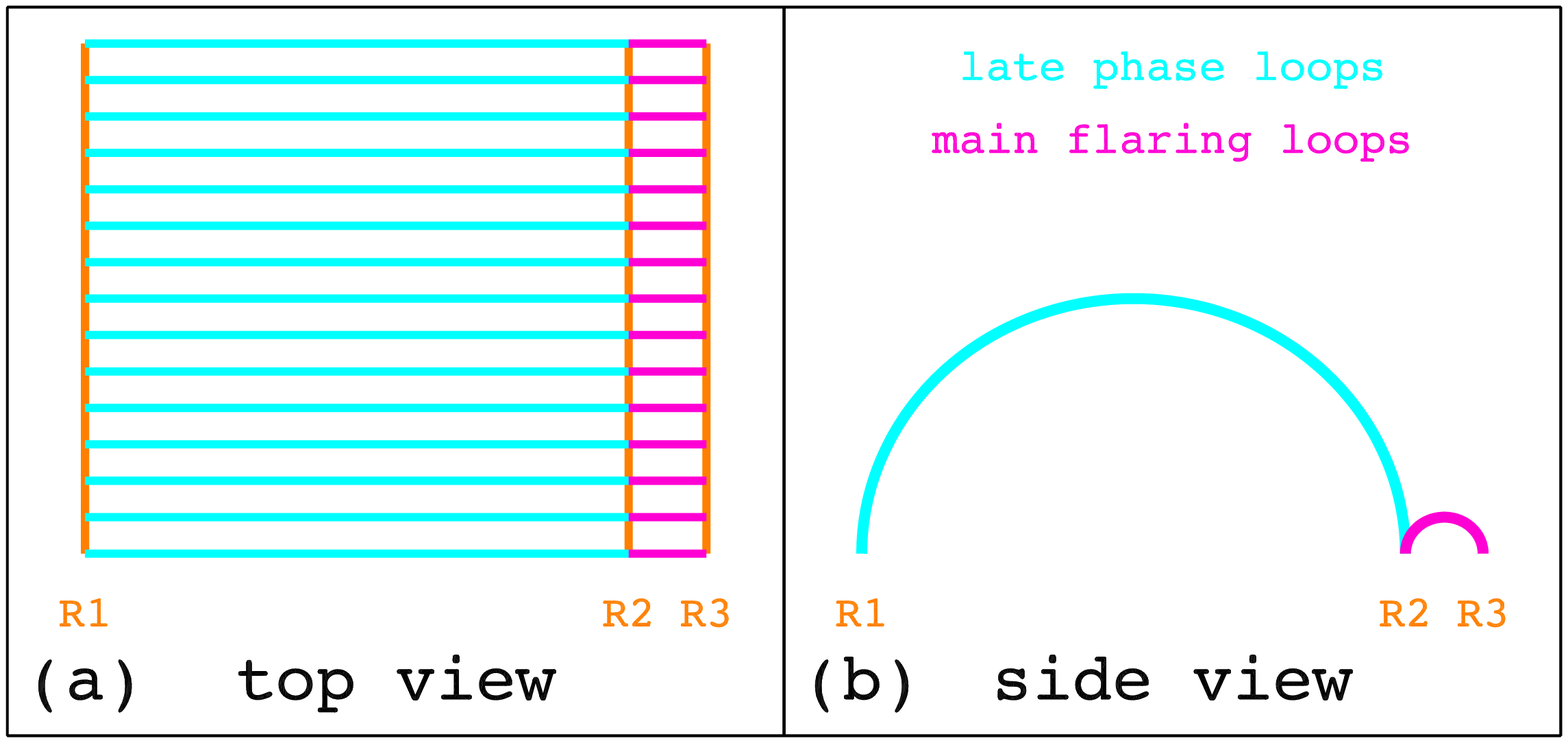}
\centering
\caption{Schematic cartoon showing the top view (a) and side view (b) of the main flaring loops (magenta lines) and late-phase loops (cyan lines), respectively.}
\label{fig5}
\end{figure}

As mentioned above, the post-flare loops are composed of two groups: the short main flaring loops and long late-phase loops. 
The former is rooted in R2 and R3, and the latter is rooted in R1 and R2.
In Fig.~\ref{fig5}, we draw a simplified schematic cartoon to illustrate the magnetic connections from top view and side view.
Both the main flaring loops (magenta lines) and late-phase loops (cyan lines) are assumed to have a semi-circular shape.
The footpoint area of main flaring loops is taken to be equal to $A_{\rm R2}$ and the footpoint area of late-phase loops is taken to be equal to $A_{\rm R1}$.
The calculated values of $A_{\rm R1}$ and $A_{\rm R2}$ are listed in the second column of Table~\ref{tab-3}.
The total length (2$L$) of the main flaring loops and late-phase loops are taken to be 10$-$25 Mm and $\sim$110 Mm based on the stereoscopic measurement \citep{dai18b}, 
which are listed in the third column of Table~\ref{tab-3}. 
Accordingly, the volume of main flaring loops is calculated to be 2$L\times A_{\rm R2}=(1.2-3.0)\times$10$^{27}$ cm$^{3}$.
The volume of late-phase loops is calculated to be 2$L\times A_{\rm R1}=1.2\times$10$^{28}$ cm$^{3}$.
The total volume of post-flare loops consisting of the main-flaring loops and late-phase loops has a range (1.3$-$1.5)$\times$10$^{28}$ cm$^{3}$.
According to Equation~\ref{eqn2}, the peak thermal energy of the flare is calculated to be (1.7$-$1.8)$\times$10$^{30}$ erg and is listed in the sixth column of Table~\ref{tab-2}.
The heating requirement of hot plasma including the peak thermal energy and radiative loss is (2.8$-$2.9)$\times$10$^{30}$ erg.

\subsection{Energy input in main flaring loops and EUV late-phase loops} \label{s-loops}
Using the EBTEL code and observed parameters, \citet{dai18b} performed a zero-dimensional (0D) numerical simulation to study the thermodynamic evolution of a typical EUV late-phase loop. 
The imposed heating rate as a function of time is expressed as:
\begin{equation} \label{eqn3}
H(t)=H_0 e^{-\frac{(t-t_0)^2}{2\tau^2}},
\end{equation}
where $H_0$ denotes the amplitude of heating rate and $\tau=23$ s denotes the Gaussian heating duration.
The best-fit value of $H_0$ (1.1 erg cm$^{-3}$ s$^{-1}$) is larger than previously reported values \citep{sun13,li14,zhou19}. 
The energy input in the flare loops is expressed as $E_{in}=V\times\int H(t) dt$ erg.
Assuming that the heating durations and amplitudes of heating rate are the same for each of the late-phase loops,
the energy input in late-phase loops reaches $\sim$7.7$\times$10$^{29}$ erg.
For the main flaring loops, as mentioned in Sect.~\ref{s-flare}, both of the two-stage energy release processes contribute to the heating.
Hence, the heating duration of the main flaring loops is double compared with the late-phase loops.
Assuming that the amplitude of heating rate is equal to $H_0$, the energy input in the main flaring loops is estimated to be (1.5$-$3.8)$\times$10$^{29}$ erg.
Based on the EBTEL simulation, the energy input in the post-flare loops amounts to (9.2$-$11.5)$\times$10$^{29}$ erg.

\begin{table*}
\centering
\caption{Physical parameters of the main flaring loops and late-phase loops, including the footpoint area, total loop length, volume, and energy input.}
\label{tab-3}
\begin{tabular}{ccccc}
\hline\hline
   & $A$           & 2$L$  & $V$          & $E_{in}$ \\ 
   & (10$^{18}$ cm$^2$)  &  (Mm) & (10$^{27}$ cm$^3$)  & (10$^{29}$ erg) \\
\hline
main flaring loops  & 1.2 & 10$-$25 & 1.2$-$3.0 & 1.5$-$3.8 \\
late-phase loops   &  1.1 & 110 & 12.1 & 7.7 \\
\hline
\end{tabular}
\end{table*}

\subsection{Nonthermal energy in flare-accelerated electrons} \label{s-nth}
\begin{figure}
\includegraphics[width=9cm,clip=]{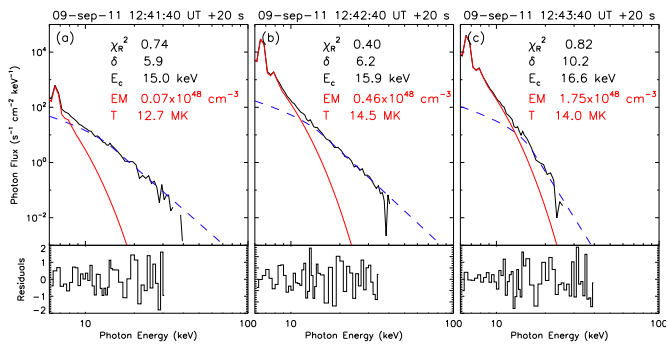}
\centering
\caption{Three characteristic HXR spectra with an integration time of 20 s (black lines) and the fitted thermal (red lines) and nonthermal (blue lines) components.
$T_{e}$ and EM of the thermal component are labeled.
The low-energy cutoff ($E_c$), spectral index ($\delta$) of nonthermal electrons, and $\chi_R^2$ of the residuals are also labeled.}
\label{fig6}
\end{figure}

Under the thick-target assumption \citep{fle13}, HXR emissions at the footpoints are produced by collisional bremsstrahlung during the passage of flare-accelerated 
nonthermal electrons through a dense target, where the electrons are stopped by Coulomb collisions and lose energy in the chromosphere. 
The localized plasma can be heated up to $\sim$10 MK accompanied by chromospheric evaporation and condensation \citep{tian14,li15,zqm16a,zqm16b,zhou20}.
The distribution of injected nonthermal electrons above a low-energy cutoff ($E_{c}$) is assumed to be in the form of single power law with a spectral index of $\delta$,
$F_0(E_0)=A_0E_0^{-\delta}$, where $A_0$ is a normalized parameter of the total electron flux (in units of 10$^{35}$ electrons s$^{-1}$).
In Fig.~\ref{fig6}, three characteristic HXR spectra made from detector 9 of RHESSI during the first stage of energy release are displayed.
In each panel, the fitted thermal component is drawn with a red line, while the nonthermal component is drawn with a blue line.
It is obvious that the isothermal temperatures from RHESSI are close to those derived from GOES.

The delivered energy by nonthermal electrons above $E_c$ is calculated by integrating the injected power during 12:40$-$12:46 UT,
\begin{equation} \label{eqn4}
  E_{nt,e}=A_{nt}\int_{t_1^\prime}^{t_2^\prime} \int_{E_c}^{E_H}E_0F_0(E_0){dE_0dt},
\end{equation}
where $A_{nt}$ is the electron injection area, $E_H$ is set to be 30 MeV \citep{war16a}.
The nonthermal energy in flare-accelerated electrons during the first stage is estimated to be $\sim$1.74$\times$10$^{30}$ erg.

Considering that the HXR observation from RHESSI was unavailable during the second stage of energy release around 12:47:30 UT, it is impossible to calculate the nonthermal 
energy of electrons from RHESSI during the second stage. In Fig.~\ref{fig2}(b), the second peak of GOES derivative accounts for $\sim$1/4 of the first peak, 
implying that energy input during the second stage accounts for $\sim$1/4 of the first stage.
In this way, the nonthermal energy in the flare-accelerated electrons during the second stage is crudely estimated to be 0.44$\times$10$^{30}$ erg.
Therefore, the total nonthermal energy in electrons lies in the range of (1.7$-$2.2)$\times$10$^{30}$ erg.
It is clear that the nonthermal energy input by flare-accelerated electrons is slightly larger than the peak thermal energy 
and is adequate to provide the energy input of main flaring loops and late-phase loops, which is consistent with the collisional thick-target model.

\subsection{Magnetic free energy} \label{s-free}
\begin{figure}
\includegraphics[width=8cm,clip=]{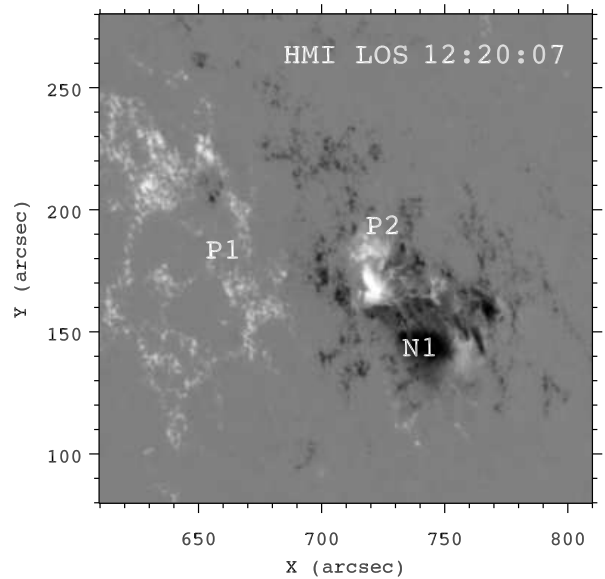}
\centering
\caption{HMI LOS magnetogram at 12:20:07 UT. P1 and P2 denote the positive polarities. N1 denotes the negative polarities surrounding P2.}
\label{fig7}
\end{figure}

As mentioned in Sect.~\ref{s-intro}, the energies in flares and/or CMEs come from the pre-stored free energy in the non-potential magnetic fields before eruption \citep{asch14}.
The magnetic energy is defined as $E_m=\int_V \frac{B^2}{8\pi}dV$. The free energy of the non-potential field is defined as $E_{mag}=E_m^{NP}-E_m^{P}$, 
where $E_m^{NP}$ and $E_m^{P}$ signify the magnetic energies in the non-potential and potential field \citep{guo08}.
Figure~\ref{fig7} shows the HMI LOS magnetogram at 12:20:07 UT, featuring a strong positive polarity (P2) surrounded by negative polarities (N1). A weak and disperse positive
polarity (P1) is located to the east of N1. The 3D magnetic topology of AR 11283 at 12:20 UT is obtained by using the NLFFF extrapolation (see Fig.\,7 of \citet{dai18b}).
The magnetic free energy before flare is calculated to be 9.1$\times$10$^{31}$ erg, which is in the same order of magnitude with the X3.4 flare on 2006 December 13 \citep{guo08}.
It is clear that the magnetic free energy is $\ge$40 times higher than the nonthermal energy in electrons.
Therefore, the free energy is sufficient to account for the plasma heating, acceleration of nonthermal particles, and radiations.

\section{Discussion} \label{s-disc}
So far, complete investigations of energy partition in eruptive flares are substantial \citep{ems04,ems05,ems12,war16a,war16b,asch17}. The energy partition in confined flares is rarely explored.
\citet{zqm19} calculated the energy components of two M1.1 confined flares in AR 12434, finding that the two flares bear some similarity in morphology, evolution, and energy partition.
Besides, the nonthermal energy in electrons is sufficient to supply the total heating requirement. 
In this paper, the M1.2 flare in AR 11283 has a much longer lifetime ($\sim$110 minutes) owing to the extremely large EUV late phase. 
The radiated energy in 1$-$70 {\AA} and total radiative loss are 4$-$6 times larger than the M1.1 CRFs. The peak thermal energy is comparable with that of CRFs.
The nonthermal energy in electrons account for $\sim$50\% that of CRFs. 

The separate radiated energies in the main phase and late phase may change from case to case. For the M1.0 confined flare on 2010 November 5, the radiation in EVE 6$-$37 nm 
in the late phase is significantly larger than that in the main phase \citep{liu15}. On the contrary, for the X1.3 eruptive flare on 2014 April 25, the radiation in the main phase is twelve times 
larger than that in the late phase \citep{zhou19}. For the M1.2 flare in our study, there is no distinct demarcation point between the main phase and late phase. 

\citet{mot20} studied a flare as a result of loop-loop interaction on 2013 November 5. One loop is tenuous and hot, while the other is dense and cool.
The released magnetic energy is converted to the nonthermal energy of accelerated electrons with a high efficiency and further converted to the thermal energy of the plasma. 
Meanwhile, the energy is deposited equally in the two flaring loops.
In our study, the main flaring loops are short (10$-$25 Mm) and the late-phase loops are long ($\sim$110 Mm).
The ratio of energy input in the main flaring loops to that in late-phase loops is $\leq$1/2 (see Table~\ref{tab-3}).

It should be emphasized that our calculation of energy components has limitations. 
The estimation of volume of the SXR-emitting plasma is based on a simplified schematic cartoon considering that the true morphology of the evolving post-flare loops is unknown.
The calculation of nonthermal energy in flare-accelerated electrons has a large uncertainty since there was no HXR observation during the second stage of energy release.
Besides, the nonthermal energy in flare-accelerated ions is not considered. The energy transported to the upper chromosphere by Alfv{\'e}n wave is not easy to quantify \citep{fle08,reep16}.
Additional case studies are desirable to get a better understanding of the complete energy partition in confined flares.

\section{Summary} \label{s-sum}
In this paper, we reanalyze the M1.2 confined flare with a large EUV late phase observed by SDO, GOES, and RHESSI on 2011 September 9, focusing on its energy partition.
Various energy components of the flare are calculated, including the magnetic free energy, peak thermal energy, nonthermal energy of flare-accelerated electrons, 
total radiative loss of hot plasma, and radiative outputs in 1$-$8 {\AA}, 1$-$70 {\AA}, and 70$-$370 {\AA}. The main results are summarized below:
\begin{enumerate}
\item{The radiation ($\sim$5.4$\times$10$^{30}$ erg) in 1$-$70 {\AA} is nearly eleven times larger than the radiation in 70$-$370 {\AA}, 
and is nearly 180 times larger than the radiation in 1$-$8 {\AA}.}
\item{The peak thermal energy of the post-flare loops is estimated to be (1.7$-$1.8)$\times$10$^{30}$ erg based on a simplified schematic cartoon.
Based on previous results of EBTEL simulation, the energy inputs in the main flaring loops and late-phase loops are (1.5$-$3.8)$\times$10$^{29}$ erg and 7.7$\times$10$^{29}$ erg, respectively.}
\item{The nonthermal energy ((1.7$-$2.2)$\times$10$^{30}$ erg) of the flare-accelerated electrons is comparable to the peak thermal energy 
and is sufficient to provide the energy input of the main flaring loops and late-phase loops.
The magnetic free energy (9.1$\times$10$^{31}$ erg) before flare is large enough to provide the heating requirement and radiation, corroborating the magnetic origin of confined flares.}
\end{enumerate}

\begin{acknowledgements}

The authors thank the referee for valuable and constructive suggestions.
SDO is a mission of NASA\rq{}s Living With a Star Program. AIA and HMI data are courtesy of the NASA/SDO science teams. 
This work is funded by the B-type Strategic Priority Program of the Chinese Academy of Sciences, Grant No. XDB41000000,
NSFC grants (No. 11773079, 11790302, 11373023), the Science and Technology Development Fund of Macau (275/2017/A),
CAS Key Laboratory of Solar Activity, National Astronomical Observatories (KLSA202006), and the Youth Innovation Promotion Association CAS.
Y.D. is also sponsored by the Open Research Project of National Center for Space Weather, China Meteorological Administration.
\end{acknowledgements}


\begin{thebibliography}{}
\bibitem[Amari et al.(2018)]{ama18} Amari, T., Canou, A., Aly, J.-J., et al.\ 2018, \nat, 554, 211
\bibitem[Aschwanden et al.(2014)]{asch14} Aschwanden, M.~J., Xu, Y., \& Jing, J.\ 2014, \apj, 797, 50
\bibitem[Aschwanden et al.(2015)]{asch15} Aschwanden, M.~J., Boerner, P., Ryan, D., et al.\ 2015, \apj, 802, 53
\bibitem[Aschwanden et al.(2017)]{asch17} Aschwanden, M.~J., Caspi, A., Cohen, C.~M.~S., et al.\ 2017, \apj, 836, 17
\bibitem[Baumgartner et al.(2018)]{bau18} Baumgartner, C., Thalmann, J.~K., \& Veronig, A.~M.\ 2018, \apj, 853, 105
\bibitem[Brown(1971)]{bro71} Brown, J.~C.\ 1971, \solphys, 18, 489
\bibitem[Cai et al.(2021)]{cai21} Cai, Z.~M., Zhang, Q.~M., Ning, Z.~J., et al.\ 2021, arXiv:2102.09819
\bibitem[Cargill(1994)]{car94} Cargill, P.~J.\ 1994, \apj, 422, 381
\bibitem[Cargill et al.(2012a)]{car12a} Cargill, P.~J., Bradshaw, S.~J., \& Klimchuk, J.~A.\ 2012a, \apj, 752, 161
\bibitem[Cargill et al.(2012b)]{car12b} Cargill, P.~J., Bradshaw, S.~J., \& Klimchuk, J.~A.\ 2012b, \apj, 758, 5
\bibitem[Chamberlin et al.(2012)]{cham12} Chamberlin, P.~C., Milligan, R.~O., \& Woods, T.~N.\ 2012, \solphys, 279, 23
\bibitem[Chen et al.(2020)]{chen20} Chen, J., Liu, R., Liu, K., et al.\ 2020, \apj, 890, 158
\bibitem[Cox \& Tucker(1969)]{cox69} Cox, D.~P. \& Tucker, W.~H.\ 1969, \apj, 157, 1157
\bibitem[Dai et al.(2013)]{dai13} Dai, Y., Ding, M.~D., \& Guo, Y.\ 2013, \apj, 773, L21
\bibitem[Dai \& Ding(2018)]{dai18a} Dai, Y., \& Ding, M.\ 2018, \apj, 857, 99
\bibitem[Dai et al.(2018)]{dai18b} Dai, Y., Ding, M., Zong, W., et al.\ 2018, \apj, 863, 124
\bibitem[Demoulin et al.(1996)]{dem96} Demoulin, P., Henoux, J.~C., Priest, E.~R., \& Mandrini, C.~H.\ 1996, \aap, 308, 643
\bibitem[Emslie et al.(2004)]{ems04} Emslie, A.~G., Kucharek, H., Dennis, B.~R., et al.\ 2004, Journal of Geophysical Research (Space Physics), 109, A10104
\bibitem[Emslie et al.(2005)]{ems05} Emslie, A.~G., Dennis, B.~R., Holman, G.~D., et al.\ 2005, Journal of Geophysical Research (Space Physics), 110, A11103
\bibitem[Emslie et al.(2012)]{ems12} Emslie, A.~G., Dennis, B.~R., Shih, A.~Y., et al.\ 2012, \apj, 759, 71
\bibitem[Feng et al.(2013)]{feng13} Feng, L., Wiegelmann, T., Su, Y., et al.\ 2013, \apj, 765, 37
\bibitem[Fletcher \& Hudson(2008)]{fle08} Fletcher, L. \& Hudson, H.~S.\ 2008, \apj, 675, 1645
\bibitem[Fletcher et al.(2011)]{fle11} Fletcher, L., Dennis, B.~R., Hudson, H.~S., et al.\ 2011, \ssr, 159, 19
\bibitem[Fletcher et al.(2013)]{fle13} Fletcher, L., Hannah, I.~G., Hudson, H.~S., et al.\ 2013, \apj, 771, 104
\bibitem[Forbes et al.(2006)]{for06} Forbes, T.~G., Linker, J.~A., Chen, J., et al.\ 2006, \ssr, 123, 251
\bibitem[Guo et al.(2008)]{guo08} Guo, Y., Ding, M.~D., Wiegelmann, T., et al.\ 2008, \apj, 679, 1629
\bibitem[Hannah \& Kontar(2013)]{han13} Hannah, I.~G., \& Kontar, E.~P.\ 2013, \aap, 553, A10
\bibitem[Inglis \& Christe(2014)]{ing14} Inglis, A.~R. \& Christe, S.\ 2014, \apj, 789, 116
\bibitem[Ji et al.(2003)]{ji03} Ji, H., Wang, H., Schmahl, E.~J., et al.\ 2003, \apjl, 595, L135
\bibitem[Jiang et al.(2013)]{jia13} Jiang, C., Feng, X., Wu, S.~T., et al.\ 2013, \apjl, 771, L30
\bibitem[Jing et al.(2009)]{jing09} Jing, J., Chen, P.~F., Wiegelmann, T., et al.\ 2009, \apj, 696, 84
\bibitem[Klimchuk et al.(2008)]{kli08} Klimchuk, J.~A., Patsourakos, S., \& Cargill, P.~J.\ 2008, \apj, 682, 1351
\bibitem[Kuhar et al.(2017)]{kuh17} Kuhar, M., Krucker, S., Hannah, I.~G., et al.\ 2017, \apj, 835, 6
\bibitem[Lemen et al.(2012)]{lem12} Lemen, J.~R., Title, A.~M., Akin, D.~J., et al.\ 2012, \solphys, 275, 17
\bibitem[Li et al.(2014)]{li14} Li, Y., Ding, M.~D., Guo, Y., et al.\ 2014, \apj, 793, 85
\bibitem[Li et al.(2015)]{li15} Li, D., Ning, Z.~J., \& Zhang, Q.~M.\ 2015, \apj, 813, 59
\bibitem[Lin et al.(2002)]{lin02} Lin, R.~P., Dennis, B.~R., Hurford, G.~J., et al.\ 2002, \solphys, 210, 3
\bibitem[Liu et al.(2013)]{liu13} Liu, K., Zhang, J., Wang, Y., et al.\ 2013, \apj, 768, 150
\bibitem[Liu et al.(2015)]{liu15} Liu, K., Wang, Y., Zhang, J., et al.\ 2015, \apj, 802, 35
\bibitem[Masson et al.(2017)]{mas17} Masson, S., Pariat, {\'E}., Valori, G., et al.\ 2017, \aap, 604, A76
\bibitem[Myers et al.(2015)]{my15} Myers, C.~E., Yamada, M., Ji, H., et al.\ 2015, \nat, 528, 526
\bibitem[Milligan et al.(2014)]{mil14} Milligan, R.~O., Kerr, G.~S., Dennis, B.~R., et al.\ 2014, \apj, 793, 70
\bibitem[Moore et al.(2001)]{moo01} Moore, R.~L., Sterling, A.~C., Hudson, H.~S., et al.\ 2001, \apj, 552, 833
\bibitem[Motorina et al.(2020)]{mot20} Motorina, G.~G., Fleishman, G.~D., \& Kontar, E.~P.\ 2020, \apj, 890, 75
\bibitem[Patsourakos et al.(2020)]{pats20} Patsourakos, S., Vourlidas, A., T{\"o}r{\"o}k, T., et al.\ 2020, \ssr, 216, 131
\bibitem[Pesnell et al.(2012)]{pes12} Pesnell, W.~D., Thompson, B.~J., \& Chamberlin, P.~C.\ 2012, \solphys, 275, 3
\bibitem[Prasad et al.(2020)]{pra20} Prasad, A., Dissauer, K., Hu, Q., et al.\ 2020, \apj, 903, 129
\bibitem[Priest \& Forbes(2002)]{pri02} Priest, E.~R. \& Forbes, T.~G.\ 2002, \aapr, 10, 313
\bibitem[Reames(2015)]{rea15} Reames, D.~V.\ 2015, \ssr, 194, 303. doi:10.1007/s11214-015-0210-7
\bibitem[Reep \& Russell(2016)]{reep16} Reep, J.~W. \& Russell, A.~J.~B.\ 2016, \apjl, 818, L20
\bibitem[Reeves et al.(2010)]{ree10} Reeves, K.~K., Linker, J.~A., Miki{\'c}, Z., et al.\ 2010, \apj, 721, 1547
\bibitem[Saint-Hilaire \& Benz(2002)]{sai02} Saint-Hilaire, P., \& Benz, A.~O.\ 2002, \solphys, 210, 287
\bibitem[Scherrer et al.(2012)]{sch12} Scherrer, P.~H., Schou, J., Bush, R.~I., et al.\ 2012, \solphys, 275, 207
\bibitem[Schrijver et al.(2008)]{sch08} Schrijver, C.~J., DeRosa, M.~L., Metcalf, T., et al.\ 2008, \apj, 675, 1637
\bibitem[Song et al.(2014)]{song14} Song, H.~Q., Zhang, J., Cheng, X., et al.\ 2014, \apj, 784, 48
\bibitem[Stoiser et al.(2007)]{sto07} Stoiser, S., Veronig, A.~M., Aurass, H., et al.\ 2007, \solphys, 246, 339
\bibitem[Sun et al.(2013)]{sun13} Sun, X., Hoeksema, J.~T., Liu, Y., et al.\ 2013, \apj, 778, 139
\bibitem[Sun et al.(2015)]{sun15} Sun, X., Bobra, M.~G., Hoeksema, J.~T., et al.\ 2015, \apj, 804, L28
\bibitem[Thalmann et al.(2015)]{tha15} Thalmann, J.~K., Su, Y., Temmer, M., et al.\ 2015, \apj, 801, L23
\bibitem[Tian et al.(2014)]{tian14} Tian, H., Li, G., Reeves, K.~K., et al.\ 2014, \apjl, 797, L14
\bibitem[Wang \& Zhang(2007)]{wang07} Wang, Y., \& Zhang, J.\ 2007, \apj, 665, 1428
\bibitem[Wang et al.(2020)]{wang20} Wang, Y., Ji, H., Warmuth, A., et al.\ 2020, \apj, 905, 126
\bibitem[Warmuth \& Mann(2016a)]{war16a} Warmuth, A., \& Mann, G.\ 2016a, \aap, 588, A115
\bibitem[Warmuth \& Mann(2016b)]{war16b} Warmuth, A., \& Mann, G.\ 2016b, \aap, 588, A116
\bibitem[Warmuth \& Mann(2020)]{war20} Warmuth, A. \& Mann, G.\ 2020, \aap, 644, A172
\bibitem[White et al.(2005)]{wh05} White, S.~M., Thomas, R.~J., \& Schwartz, R.~A.\ 2005, \solphys, 227, 231
\bibitem[Woods et al.(2011)]{wood11} Woods, T.~N., Hock, R., Eparvier, F., et al.\ 2011, \apj, 739, 59
\bibitem[Woods et al.(2012)]{wood12} Woods, T.~N., Eparvier, F.~G., Hock, R., et al.\ 2012, \solphys, 275, 115
\bibitem[Woods et al.(2017)]{wood17} Woods, T.~N., Caspi, A., Chamberlin, P.~C., et al.\ 2017, \apj, 835, 122
\bibitem[Yan et al.(2020)]{yan20} Yan, X., Xue, Z., Cheng, X., et al.\ 2020, \apj, 889, 106
\bibitem[Zhang et al.(2001)]{zj01} Zhang, J., Dere, K.~P., Howard, R.~A., et al.\ 2001, \apj, 559, 452
\bibitem[Zhang et al.(2015)]{zqm15} Zhang, Q.~M., Ning, Z.~J., Guo, Y., et al.\ 2015, \apj, 805, 4
\bibitem[Zhang et al.(2016a)]{zqm16a} Zhang, Q.~M., Li, D., Ning, Z.~J., et al.\ 2016a, \apj, 827, 27
\bibitem[Zhang et al.(2016b)]{zqm16b} Zhang, Q.~M., Li, D., \& Ning, Z.~J.\ 2016b, \apj, 832, 65
\bibitem[Zhang et al.(2019)]{zqm19} Zhang, Q.~M., Cheng, J.~X., Feng, L., et al.\ 2019, \apj, 883, 124
\bibitem[Zhou et al.(2019)]{zhou19} Zhou, Z., Cheng, X., Liu, L., et al.\ 2019, \apj, 878, 46
\bibitem[Zhou et al.(2020)]{zhou20} Zhou, Y.-A., Li, Y., Ding, M.~D., et al.\ 2020, \apj, 904, 95
\end{thebibliography}
\end{document}